\documentclass[twocolumn,superscriptaddress,aps,prl]{revtex4-1}
\usepackage{amssymb,amsmath}
\usepackage{graphicx}
\usepackage[colorlinks,linkcolor=blue,citecolor=blue,urlcolor=blue]{hyperref}
\hypersetup{pdfstartview={XYZ null null 1.15}}

\begin{document}

\title{Collective modes at a disordered quantum phase transition}
\author{Martin Puschmann}
\affiliation{Department of Physics, Missouri University of Science and Technology, Rolla, MO 65409, USA}
\author{Jack Crewse}
\affiliation{Department of Physics, Missouri University of Science and Technology, Rolla, MO 65409, USA}
\author{Jos\'e A. Hoyos}
\affiliation{Instituto de F\'{i}sica de S\~ao Carlos, Universidade de S\~ao Paulo,
C.P. 369, S\~ao Carlos, S\~ao Paulo 13560-970, Brazil}
\author{Thomas Vojta}
\affiliation{Department of Physics, Missouri University of Science and Technology, Rolla, MO 65409, USA}

\begin{abstract}
We study the collective excitations, i.e., the Goldstone (phase) mode and the Higgs (amplitude) mode, near the superfluid--Mott glass
quantum phase transition in a two-dimensional system of disordered bosons. Using Monte Carlo simulations
as well as an inhomogeneous quantum mean-field theory with Gaussian fluctuations, we show that the Higgs mode is
strongly localized for all energies, leading to a noncritical scalar response. In contrast, the lowest-energy
Goldstone mode undergoes a striking delocalization transition as the system enters the superfluid phase. We discuss the generality of
these findings and experimental consequences, and we point out potential relations to many-body localization.
\end{abstract}

\date{\today}

\maketitle


Understanding the rich behavior that arises when many quantum particles interact with each other
remains one of the major challenges of modern condensed matter physics. Zero-temperature phase
transitions between different quantum ground states have emerged as a central ordering principle
in this field \cite{Sachdev_book11,SGCS97,Vojta_review00,VojtaM03,LRVW07,SiSteglich10}.
These quantum phase transitions (QPTs) control large regions of a material's phase diagram and lead to
unconventional thermodynamic and transport properties. Moreover, fluctuations associated with
these transitions can induce novel phases, increasing the complexity of quantum matter.

Since impurities, defects, and other types of quenched disorder are unavoidable in most
condensed matter systems, the effects of randomness on QPTs have been
studied intensively over the last two decades, leading to the discovery of exotic phenomena
such as infinite-randomness critical points \cite{Fisher92,*Fisher95}, smeared phase transitions
\cite{Vojta03a,*HoyosVojta08}, and quantum Griffiths singularities \cite{ThillHuse95,*RiegerYoung96}.
Today, the thermodynamics of many disordered QPTs is well understood,
and classification schemes \cite{MMHF00,VojtaSchmalian05,*VojtaHoyos14}
have been established based on the scaling of the disorder
strength under coarse graining as well as on the importance of rare disorder fluctuations
(see, e.g., Ref. \cite{Vojta06,*Vojta10,*Vojta19} and references therein).

Much less is known about the character and dynamics of excitations at disordered QPTs
even though excitations are crucial for a host of experiments
ranging from inelastic neutron scattering in magnetic materials to various electrical and
thermal transport measurements. Of particular interest are the collective excitations
that emerge in systems with spontaneously broken continuous symmetries. These include one or more Goldstone modes
that are related to oscillations of the order parameter direction and an amplitude (Higgs) mode that is
related to oscillations of the order parameter magnitude. Examples of such modes can be found
in superfluids, superconductors, incommensurate charge density waves, as well as planar and Heisenberg
magnets (see, e.g., Refs.\ \cite{Burgess00,PekkerVarma15}).

In this Letter, we therefore investigate the excitations close to a paradigmatic disordered QPT,
the superfluid-Mott glass transition of disordered bosons, by means of
Monte Carlo simulations and an inhomogeneous mean-field theory with Gaussian fluctuations. Our results can be
summarized as follows. Even though the thermodynamic critical behavior of the superfluid-Mott glass transition
is of conventional power-law type \cite{ProkofevSvistunov04,Vojtaetal16}, the Higgs and Goldstone modes
feature unconventional dynamics that violates naive scaling. Specifically, the Higgs mode is strongly
localized, resulting in a broad, noncritical spectral density close to the QPT.
In contrast, the incipient Goldstone mode features a striking delocalization transition as the system
enters the superfluid phase, irrespective of the disorder strength.

In the remainder of this Letter, we first introduce our model and then discuss the Monte Carlo simulations.
To explain the unusual, noncritical response observed in these simulations, we study Gaussian fluctuations
about an inhomogeneous quantum mean-field theory. We also
discuss possible experiments, and consider relations to many-body localization.

We start from a square-lattice Bose-Hubbard Hamiltonian
\begin{eqnarray}
H &=& \frac 1 2  \sum_i U_i (n_i - \bar n)^2 -\sum_{\langle ij \rangle} J_{ij} (a_i^\dagger a_j + \textrm{h.c.})
\label{eq:Hubbard}
\end{eqnarray}
with large integer filling $\bar n$. Here $a_i^\dagger$ and $a$ are the boson creation and annihilation operators
at site $i$, and $n_i=a_i^\dagger a_i$ is the number operator. If the interactions $U_i$ and the nearest-neighbor
hopping terms $J_{ij}$ are spatially uniform, the system undergoes a QPT between a superfluid ground state
(for $J \gg U$) and a gapped, incompressible Mott insulator (for $J \ll U$). In the presence of quenched disorder,
these two bulk phases are separated by the Mott glass phase, a gapless but incompressible insulator
\cite{GiamarchiLeDoussalOrignac01,WeichmanMukhopadhyay08}. In the following, we introduce
the disorder via site dilution, i.e., we randomly remove a nonzero fraction $p$ of lattice sites while
the $U_i$ and $J_{ij}$ of the remaining sites stay uniform.

To study the collective modes across the superfluid-Mott glass transition, we map the
Bose-Hubbard model (\ref{eq:Hubbard}) onto a $(2+1)$-dimensional XY model \cite{WSGY94} with columnar defects.
We then perform large-scale Monte Carlo simulations
for lattices with linear sizes of up to $L=256$ and $L_\tau=512$ in the space and imaginary time directions.
The phase diagram and the thermodynamic critical behavior (which is of conventional power-law type)
are known accurately from earlier studies
\cite{Vojtaetal16,LerchVojta19}.
For details of the simulations and the data analysis see the Supplemental Material \footnote{See Supplemental Material at XXX with Refs.\
\cite{Wolff89,MRRT53,GuoBhattHuse94,RiegerYoung94,SknepnekVojtaVojta04,*VojtaSknepnek06,BFMM98,ZWNHV15,CCFS86,HansenOleary93,BergeronTremblay16,MacKinnonKramer83,MacKinnon85}
for details of the Monte Carlo simulations, the scaling form of the scalar susceptibility, the maximum-entropy method, and
the quantum mean-field theory.}.

To analyze the Higgs mode, we compute the (disorder-averaged) imaginary-time scalar susceptibility,
\begin{eqnarray}
\chi_{\rho\rho}(\mathbf x, \tau) &=&   \langle \rho(\mathbf x,\tau)\rho(0,0) \rangle
                              -\langle \rho(\mathbf x,\tau)\rangle \langle \rho(0,0) \rangle ~
\end{eqnarray}
and its Fourier transform $\tilde \chi_{\rho\rho}(\mathbf q, i\omega_m)$. Here, $\rho(\mathbf x,\tau)$ is the
local order parameter amplitude, obtained as the average of the XY variables over a small (five-site) cluster.
The dynamic susceptibility is given by the analytic continuation from imaginary Matsubara frequencies $i \omega_m$ to real frequencies $\omega$,
\begin{eqnarray}
\chi_{\rho\rho}(\mathbf q, \omega)  = \tilde \chi_{\rho\rho}(\mathbf q, i\omega_m \to \omega +i0^+)~.
\end{eqnarray}
Unfortunately, the analytic continuation is an ill-posed problem and sensitive to Monte Carlo
noise. To overcome this problem, we employ a maximum-entropy (MaxEnt) method
\cite{JarrellGubernatis96}. Its technical details and robustness are discussed in
the Suppl.\ Material \cite{Note1}.
Generalizing scaling arguments of Podolsky and Sachdev \cite{PodolskySachdev12} to the disordered case
suggests that the singular part of the scalar susceptibility in $d$ space dimensions has the form
\begin{eqnarray}
\chi_{\rho\rho}(\mathbf q, \omega)  = \omega^{ [(d+z)\nu - 2]/(\nu z) } X (\mathbf q r^{-\nu}, \omega r^{-\nu z})
\label{eq:chirhorho_scaling}
\end{eqnarray}
where $r$ is the distance from criticality, $\nu$ is the correlation length exponent, $z$ is the dynamical exponent, and
$X$ is a universal scaling function \cite{Note1}.

\begin{figure}
\includegraphics[width=8.5cm]{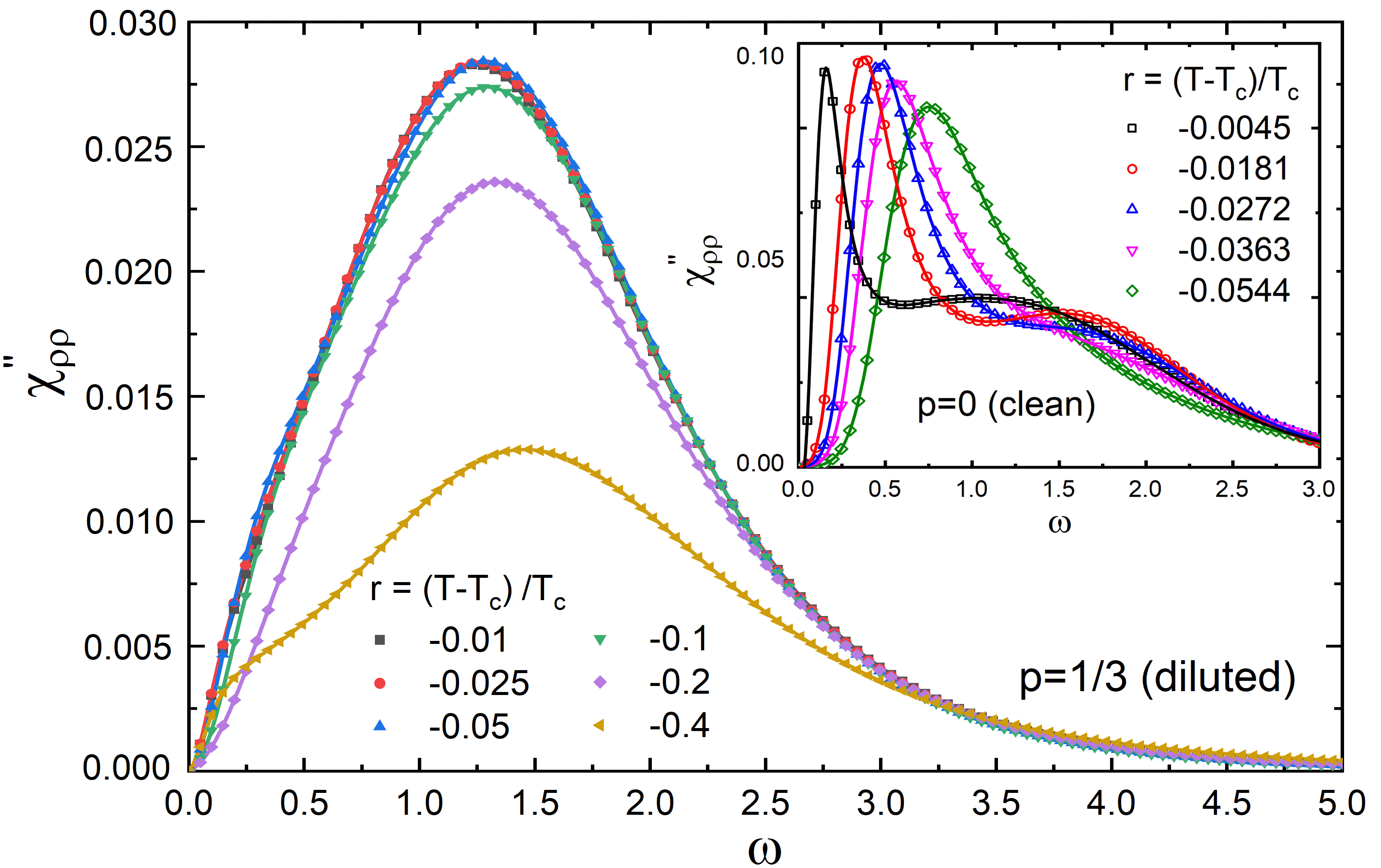}
\caption{Spectral function $\chi_{\rho\rho}''(\mathbf q=0, \omega)$ for different distances $r$ from criticality
on the superfluid side of the transition.
Main panel: dilution $p=1/3$, results averaged over 10,000 samples of sizes
$L=100, L_\tau=452$. Inset: clean case ($p=0$), $L=L_\tau=128$.
Statistical errors are small, about one symbol size;
variations of the MaxEnt parameters can shift
the peak positions systematically
by up to about 10\% \cite{Note1}.
$T$ is the Monte Carlo temperature,
not the physical temperature of the Bose-Hubbard Hamiltonian.
}
\label{fig:MC_scalar_susc}
\end{figure}

We now turn to the results of the Monte Carlo simulations. Figure \ref{fig:MC_scalar_susc} shows the spectral function
$\chi_{\rho\rho}''(\mathbf q, \omega)$ at $\mathbf q=0$ on superfluid side of the QPT,
contrasting the clean case ($p=0$) with a diluted case ($p=1/3$).
The clean spectral function features a pronounced low-energy Higgs peak that softens as the
transition is approached. The low-energy part of $\chi_{\rho\rho}''$ fulfills the scaling form
(\ref{eq:chirhorho_scaling}) in good approximation, using the exponents $\nu=0.671$ and $z=1$
of the clean 3d XY universality class \cite{CHPV06} (see Fig.\ S1 in the Suppl.\ Material \cite{Note1}).
These findings agree with previous simulations of the Higgs mode at the clean superfluid-Mott
insulator transition \cite{GazitPodolskyAuerbach13,Chenetal13}.

The spectral function of the diluted system behaves
very differently. Instead of a narrow low-energy peak, it features a broad maximum at higher energies. Importantly,
the position of this maximum is only weakly dependent on the distance $r$ from criticality; it does not vanish
for $r \to 0$. This behavior violates the scaling form (\ref{eq:chirhorho_scaling}), implying that the scalar
susceptibility is dominated by a noncritical contribution.

\begin{figure}
\includegraphics[width=8.5cm]{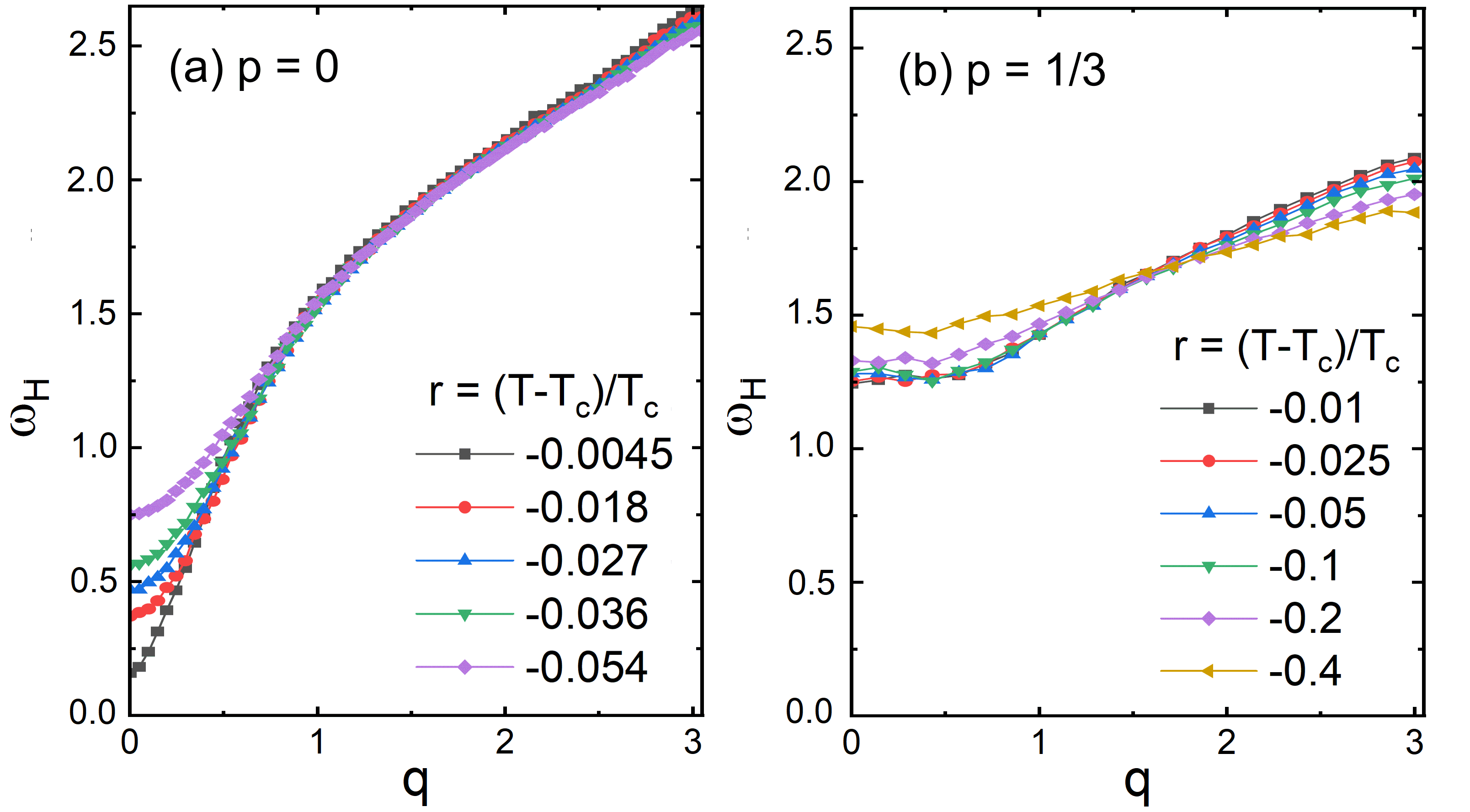}
\caption{Peak position $\omega_H$ of the spectral function $\chi_{\rho\rho}''(\mathbf q, \omega)$ vs.\
wave vector $|\mathbf q|$ (along the coordinate directions) for different distances $r$ from criticality.
(a) dilution $p=0$. (b) $p=1/3$. The simulation parameters agree with
Fig.\  \ref{fig:MC_scalar_susc}. Statistical errors are about a symbol size or less.}
\label{fig:MC_dispersion}
\end{figure}

We also study the dispersion $\omega_H(\mathbf q)$ of the peak position as a function of
the wave vector $\mathbf q$; the results are presented in Fig.\ \ref{fig:MC_dispersion}.
In the clean case, the data show the behavior expected for a $z=1$ quantum critical point. The low-energy dispersion
is linear, $\omega_H \sim |\mathbf q|$, at criticality. As $r$ increases, it crosses over to the quadratic form
$\omega_H = m_H +c \mathbf q^2$. In contrast, the dispersion of the diluted system does not change much with
the distance from criticality, and the peak energy $\omega_H$ is almost independent of $\mathbf q$ for small wave vectors.

What causes the broad, uncritical scalar response near the superfluid-Mott glass transition?
Potential reasons include increased damping and localization effects. To gain further insight and to disentangle
these possibilities, we complement the Monte Carlo simulations by an inhomogeneous mean-field theory with Gaussian fluctuations.
Our approach generalizes the theories of Refs.\ \cite{AltmanAuerbach02,Pekkeretal12} to the disordered
case. It is also related to the bond-operator method for disordered magnets \cite{VojtaM13}.

Close to the Mott phase, particle number fluctuations are small. We thus truncate the local Hilbert space
at site $j$ to three basis states, $|-_j\rangle,  |0_j\rangle$, and $|+_j\rangle$, corresponding to the
boson numbers $n_j = \bar n -1, \bar n$, and $\bar n +1$, respectively. The mean-field theory derives
from  the variational ground state wave function $|\Phi_0 \rangle = \prod_j |\phi_{0j}\rangle$ with
\begin{eqnarray}
|\phi_{0j}\rangle &=& \cos(\theta_j/ 2) |0_j\rangle  \nonumber \\
                   && + \sin(\theta_j/ 2) \left( e^{i \eta_j} |+_j\rangle + e^{-i \eta_j} |-_j\rangle \right)/\sqrt 2~.
\end{eqnarray}
It captures both the Mott state, $\theta_j =0$, and the superfluid state, $\theta_j>0$,
with the local superfluid order parameter $\langle a_j^\dagger \rangle \propto  \psi_j = \sin \theta_j e^{-i \eta_j}$.

The variational ground state energy $E_0 = \langle \Phi_0 | H | \Phi_0 \rangle$ is minimized
by uniform phases $\eta_j=\eta=\textrm{const}$ (which we set to zero in the following)
and mixing angles $\theta_i$ that fulfill the mean-field
equations
\begin{equation}
U_i \sin \theta_i - 4 \bar n \cos \theta_i \sum_j J_{ij} \sin \theta_j = 0~.
\label{eq:MF}
\end{equation}

To describe excitations on top of the mean-field ground state,
we rotate the basis in the three-state local Hilbert space of site $j$
to $|\phi_{0j}\rangle, |\phi_{Hj}\rangle, |\phi_{Gj}\rangle$
where
\begin{eqnarray}
|\phi_{Hj}\rangle &=& \sin(\theta_j/ 2) |0_j\rangle - \cos (\theta_j / 2) \left(  |+_j\rangle + |-_j\rangle \right)/\sqrt{2} \quad \nonumber \\
|\phi_{Gj}\rangle &=&  i \left(  |+_j\rangle - |-_j\rangle \right)/\sqrt{2}
\end{eqnarray}
are related to changes of order parameter magnitude and phase, respectively, compared to $|\phi_{0j}\rangle$.
The boson operators $b^\dagger_{0j}, b^\dagger_{Hj}$, and $b^\dagger_{Gj}$ create these states out of the fictitious vacuum
and fulfill the local constraint $\sum_\alpha b^\dagger_{\alpha j} b_{\alpha j} = 1$.
We now rewrite the Hamiltonian (\ref{eq:Hubbard}) in terms of the $b$ bosons, using the constraint to eliminate
(``fully condense'') $b_{0j}$ such that $b^\dagger_{Hj}$ and $b^\dagger_{Gj}$
create excitations on top of the mean-field ground state. To quadratic (Gaussian) order in $b$, the Hamiltonian decouples
into Higgs and Goldstone parts, $H= E_0 + H_H + H_G$, which both take the form
\begin{eqnarray}
H_\alpha= \sum_i A_{\alpha i} b^\dagger_{\alpha i} b_{\alpha i} + \sum_{\langle ij \rangle} B_{\alpha ij} (b^\dagger_{\alpha i} + b_{\alpha i})(b^\dagger_{\alpha j} + b_{\alpha j})~,~
\label{eq:H_H}
\end{eqnarray}
($\alpha=H,G$). The coefficients $A_{\alpha i}$ and $B_{\alpha ij}$ are nonuniform and depend on the
mixing angles $\theta_i$.
$H_H$ and $H_G$ can be diagonalized numerically by bosonic Bogoliubov transformations,
$b_{\alpha j} = \sum_k ( u_{\alpha jk} d_{\alpha k} + v_{\alpha jk}^\ast d_{\alpha k}^\dagger )$ \cite{Note1}.

We now present the results of the mean-field theory.
In the absence of dilution, $p=0$, the mean-field equation (\ref{eq:MF}) can be solved analytically.
A superfluid solution appears for $U < U_c^0 = 4 \bar n J z$ where $z=4$ is the coordination number of the lattice;
it has a uniform mixing angle $\cos \theta =U/U_c^0$ and order parameter $\psi= (1-U/U_c^0)^{1/2}$.
As the system is translationally invariant, all excitations have plane wave character. In the superfluid phase,
the Goldstone mode is gapless while the gapped Higgs mode softens at the QPT. In the insulating
phase the two modes are gapped and degenerate. All clean mean-field results agree with earlier work \cite{AltmanAuerbach02}.

\begin{figure}
\includegraphics[width=8.5cm]{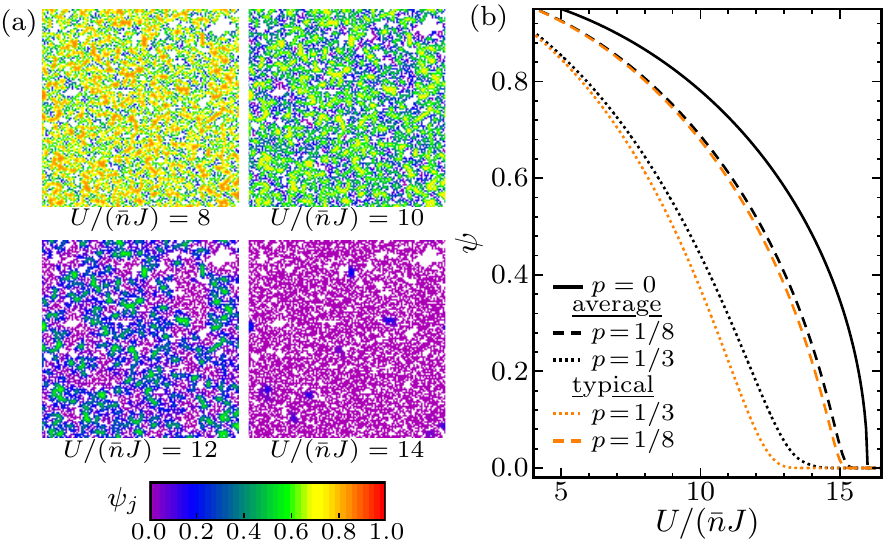}
\caption{(a) Local order parameter $\psi_j$ for several $U/(\bar n J)$ for for a system of $128^2$ sites with dilution $p=1/3$. (b)
Average and typical (geometric average) local order parameter $\psi$ as function of $U/(\bar n J)$
for dilutions $p=0, 1/8$, and $1/3$, using 1000 disorder realizations. Statistical errors are comparable to the line widths.}
\label{fig:OP_p03333}
\end{figure}

The behavior changes dramatically in the presence of disorder. Figure \ref{fig:OP_p03333} shows
the average and typical order parameter for site dilutions $p=1/8$ and $1/3$,
resulting from a numerical solution of the mean-field equations (\ref{eq:MF})
\footnote{For each lattice, we only consider the infinite percolation cluster as finite clusters cannot support
superfluid long-range order.}.
As expected, the onset of superfluidity is suppressed compared to the clean case, $p=0$. The large difference
between the average and typical order parameter for $U/(\bar n J)$ slightly below the onset of superfluidity
indicates the coexistence of superfluid puddles with insulating regions, characteristic of a Griffiths phase
(which is wider for stronger dilution). At lower $U$, the order parameter is only moderately inhomogeneous.

\begin{figure}
\includegraphics[width=8.5cm]{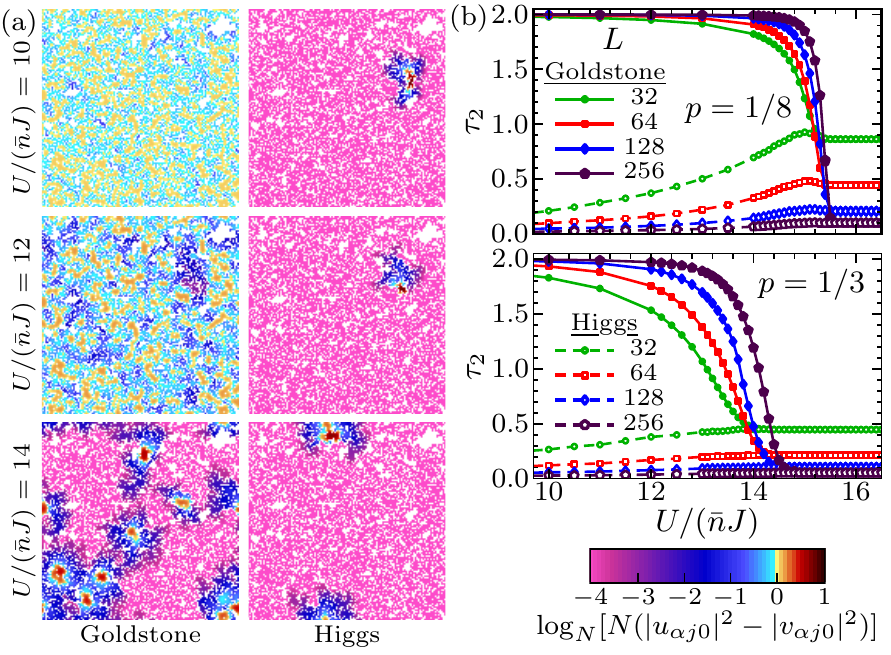}
\caption{(a) Wave functions of the lowest-energy Goldstone and Higgs modes for $p=1/3$ and several $U/(\bar n J)$,
visualized as $|u_{\alpha j0}|^2 - |v_{\alpha j0}|^2$.
(b) Generalized dimension $\tau_2$ of the lowest-energy Goldstone and Higgs modes
vs.\ interaction $U/(\bar n J)$ for $p=1/8$ and $1/3$
(averaged over 1000 disorder realizations). Statistical errors are smaller
than the symbol size.}
\label{fig:eigenmodes}
\end{figure}

Turning to excitations on top of the  mean-field ground state, Fig.\ \ref{fig:eigenmodes}a visualizes
examples of the lowest-energy eigenstates in both the Higgs and Goldstone channels for dilution $p=1/3$.
Clearly, these states show nontrivial localization properties. To characterize the localization,
we calculate the inverse participation number $P^{-1}(0)=\sum_j (|u_{\alpha j0}|^2 - |v_{\alpha j0}|^2)^2$ \cite{VojtaM13}
and the corresponding generalized dimension $\tau_2(0) =   \ln P(0) / \ln L$
\footnote{In our numerical calculations, we compute $\tau_2$ via the box-counting method,
see the Suppl.\ Material \cite{Note1}.}.
The dependence of $\tau_2$ on the interaction $U$ for the lowest-energy eigenstates
in the Higgs and Goldstone channels is presented in Fig.\ \ref{fig:eigenmodes}b. For both
weak and strong dilutions, $p=1/8$ and 1/3, we observe the same behavior.
In the insulating phase, both excitations are degenerate and strongly
localized as indicated by the rapid drop of $\tau_2$ towards zero with increasing $L$.

Upon entering the superfluid phase with decreasing $U$, the two excitations evolve
in opposite direction. The Higgs mode becomes even more localized, reflected in a further
decrease of $\tau_2$.
In contrast, the lowest Goldstone excitation undergoes a rapid delocalization transition.
Its dimension $\tau_2$ increases quickly, and its $L$-dependence
changes sign. It now increases towards $\tau_2=2$ with increasing $L$, indicating an
extended state. Within our numerical accuracy, the crossing of the $\tau_2$ vs.\ $U/(\bar n J)$
curves coincides with the onset of superfluid order. In fact, we have derived
an analytic expression for the wave function of the lowest Goldstone excitation in
the superfluid phase that proves that it is extended whenever the system features
a macroscopic order parameter \cite{Note1}.

We also study the dependence of the localization on the excitation energy \cite{Note1}.
On the insulating side, the excitations are strongly localized for all energies, and the
same is true for the Higgs mode in the superfluid phase. Goldstone excitations with nonzero energy
appear to be localized as well, with a localization length that diverges with vanishing energy.
We do not find any evidence for a mobility edge at nonzero energy, in contrast to
the Bose glass results reported in Ref.\ \cite{ZunigaLaflorencie13}.

\begin{figure}
\includegraphics[width=8.5cm]{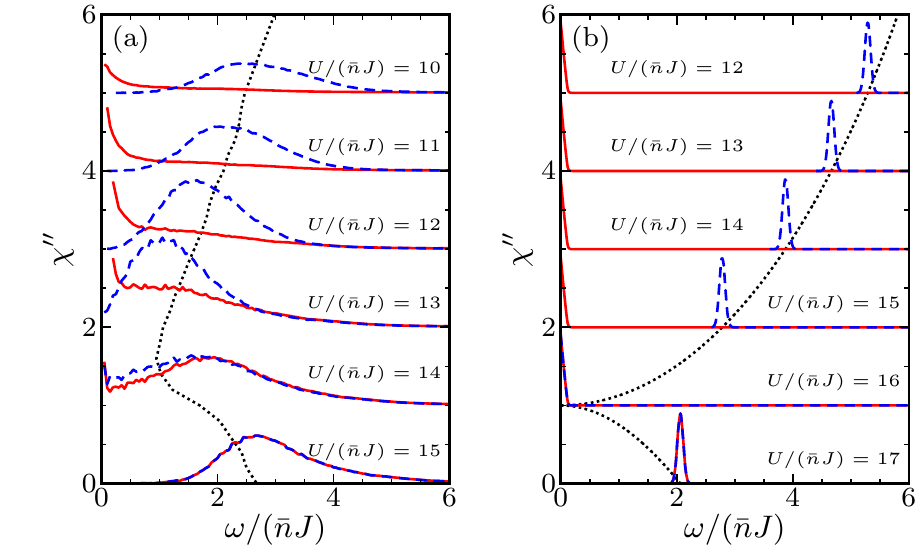}
\caption{Spectral functions $\chi''(\mathbf q = 0, \omega)$ of the Goldstone (solid lines) Higgs (dashed lines) excitations
for several interactions $U/(\bar n J)$. The curves are shifted upwards with increasing $U$. Dotted lines mark the
position of the Higgs peak in $\chi''$. (a) Dilution $p=1/3$ (240 disorder realizations, statistical errors are comparable to the line widths). (b) Clean case, $p=0$; here the peaks in the figure represent $\delta$ functions.
}
\label{fig:spectral_functions}
\end{figure}

To establish a connection to the Monte Carlo simulations, we compute the spectral densities
of the Higgs and Goldstone Green functions $\chi_{\alpha jk}(t)= -i\Theta(t) \langle
[b_{\alpha j}^\dagger(t) +b_{\alpha j}(t) , b_{\alpha k}^\dagger(0) +b_{\alpha k}(0)] \rangle$
with $\alpha=G,H$.
Figure \ref{fig:spectral_functions} shows the spectral densities
at zero wave vector for several interactions $U/(\bar n J)$, comparing the clean case with
dilution $p=1/3$.
The spectral densities of the diluted system are very broad, even though the eigenmodes are noninteracting within
the Gaussian approximation and thus have no intrinsic width. This demonstrates that the broadening of $\chi''$ is due to disorder-induced localization effects. Moreover, the peak in the Higgs spectral function does not soften
at the superfluid-Mott glass transition, mirroring the Monte Carlo results in Fig.\ \ref{fig:MC_scalar_susc}.
In contrast, the clean spectral functions show the expected $\delta$ peaks
at energies corresponding to the Higgs and Goldstone masses.

To summarize, we found the Higgs mode to be strongly localized across the superfluid-Mott glass QPT;
the scalar response is thus noncritical and violates naive scaling.
The lowest Goldstone excitation, in contrast, delocalizes upon entering the superfluid phase. Higher-energy Goldstone excitations
are localized, implying the absence of a nonzero-energy mobility edge for the excitations.
%

The mean-field theory used in the second half of this Letter provides only an approximate description of the
superfluid-Mott glass transition. In particular, it does not correctly capture rare regions effects because it cannot
describe the fluctuations of large superfluid puddles in an insulating matrix.
Whereas rare regions are known to be unimportant for the thermodynamics of this
QPT \cite{Vojta06,*Vojta10,*Vojta19}, their effects on excitations are less well understood.
Moreover, the Gaussian approximation for $H_H$ and $H_G$ neglects anharmonic effects (which could be included
by keeping higher-order terms in the expansion of $H$).
However, the agreement between the mean-field results and the numerically exact Monte Carlo simulations
gives us confidence in their validity.

Potential routes to analyze the superfluid-Mott glass transition experimentally include ultracold atoms,
dirty and granular superconductors, as well as diluted quantum antiferromagnets.
Recently, the effects of the Higgs mode on the dynamical conductivity in disordered superconducting
thin films were modeled by a bosonic Hamiltonian similar to ours \cite{SLRT14,Shermanetal15}. The Monte Carlo data in these
papers appear to be compatible with a more conventional scenario in which the Higgs response sharpens and softens as the
QPT is approached. We believe that this may stem from the comparatively weak disorder used in Refs.\
\cite{SLRT14,Shermanetal15} which causes a  slow crossover to the disordered behavior
\footnote{This is supported by the fact that the critical behavior found in Ref.\ \cite{SLRT14} agreed
with a clean dynamical exponent $z=1$ rather than the disordered value $z=1.52$ \cite{Vojtaetal16}.
}.

In conclusion, our work demonstrates that disordered QPTs can feature unconventional
collective excitations even if their thermodynamic critical behavior is completely regular.
This implies a number of important general questions about collective modes at disordered QPTs:
Can one classify the excitation dynamics along similar lines as the thermodynamics?
What is the character (and critical behavior) of the delocalization
transition of the Goldstone mode? Under what conditions does a mobility edge appear? Is it related to many-body localization?
What role is played by the space dimensionality? These questions remain tasks for the future.

\begin{acknowledgments}
This work was supported by the NSF under Grant Nos.\ DMR-1506152, DMR-1828489, PHY-1125915 and PHY-1607611,
by Conselho Nacional de Desenvolvimento Cient\'{\i}fico e Tecnol\'{o}gico (CNPq) under Grant No. 312352/2018-2, and by
FAPESP under Grants No. 2015/23849-7 and No. 2016/10826-1.
T.V. and J.A.H. acknowledge the hospitality of the Aspen Center for Physics, and T.V. thanks the
Kavli Institute for Theoretical Physics where part of the work was performed.
\end{acknowledgments}

%

\clearpage       

\setcounter{equation}{0}
\setcounter{figure}{0}
\setcounter{table}{0}
\setcounter{page}{1}
\makeatletter
\renewcommand{\theequation}{S\arabic{equation}}
\renewcommand{\theHequation}{S\arabic{equation}}
\renewcommand{\thefigure}{S\arabic{figure}}
\renewcommand{\theHfigure}{S\arabic{figure}}
\renewcommand{\bibnumfmt}[1]{[S#1]}
\renewcommand{\citenumfont}[1]{S#1}

\onecolumngrid
\begin{center}
{\large\bf Collective modes at a disordered quantum phase transition: Supplemental material}

\bigskip

Martin Puschmann$^1$, Jack Crewse$^1$, Jos\'e A. Hoyos$^2$, and Thomas Vojta $^1$

\smallskip

{\it
$^1$Department of Physics, Missouri University of Science and Technology, Rolla, MO 65409, USA\\
$^2$Instituto de F\'{i}sica de S\~ao Carlos, Universidade de S\~ao Paulo,\\
C.P. 369, S\~ao Carlos, S\~ao Paulo 13560-970, Brazil
}\\
(Dated: \today)

\vspace*{5mm}
\end{center}
\twocolumngrid

In the following sections, we provide technical details about the Monte Carlo simulations,
the scaling form of the scalar susceptibility,
the maximum-entropy method used to analytically continue the imaginary-frequency susceptibilities,
and the quantum mean-field theory.

\section*{S1. Details of the Monte Carlo simulations}

The Monte Carlo simulations follow the approach used in Refs.\ \cite{Vojtaetal16S,LerchVojta19S}
to study the thermodynamic critical behavior. For large integer filling $\bar n$, the
square-lattice Bose-Hubbard Hamiltonian
\begin{eqnarray}
H &=& \frac 1 2  \sum_i U_i (n_i - \bar n)^2 -\sum_{\langle ij \rangle} J_{ij} (a_i^\dagger a_j + \textrm{h.c.})~
\label{eq:S_Hubbard}
\end{eqnarray}
can be mapped \cite{WSGY94S} onto a classical $(2+1)$-dimensional XY model on a cubic lattice.
If the disorder is introduced by means of site dilution, the resulting classical Hamiltonian reads
\begin{equation}
 H_{\rm cl} = -J_s\sum_{\langle i,j\rangle, \tau}
\epsilon_i\epsilon_j\mathbf{S}_{i,\tau}\cdot\mathbf{S}_{j,\tau}
-J_t \sum_{i,\tau}\epsilon_i\mathbf{S}_{i,\tau}\cdot\mathbf{S}_{i,\tau+1}
\label{eq:S_Hcl}
\end{equation}
where $\mathbf{S}_{i,\tau}$ is a two-component unit vector at the lattice site with spatial
coordinate $i$ and ``imaginary-time'' coordinate $\tau$.
The independent quenched random variables $\epsilon_i$
take the values 0 (vacancy) with probability $p$ and 1 (occupied site) with probability $1-p$.
Because the vacancy positions do not depend on the imaginary-time coordinate $\tau$, the
defects in the classical model (\ref{eq:S_Hcl}) are columnar.
The values of the coupling constants $J_s/T$ and $J_t/T$ depend on the parameters of the original
Bose-Hubbard model. $T$ is the ``classical'' temperature of the Hamiltonian (\ref{eq:S_Hcl})
whereas the physical temperature
of the Bose-Hubbard model (\ref{eq:S_Hubbard}) maps onto the inverse system size in imaginary-time direction of
the classical model.
As we are interested in universal properties, the values of $J_s$ and $J_t$ are not important
for the qualitative behavior. We therefore set $J_s=J_t=1$ and vary the classical temperature $T$ to tune through the
transition.

\begin{figure}
\includegraphics[width=\columnwidth]{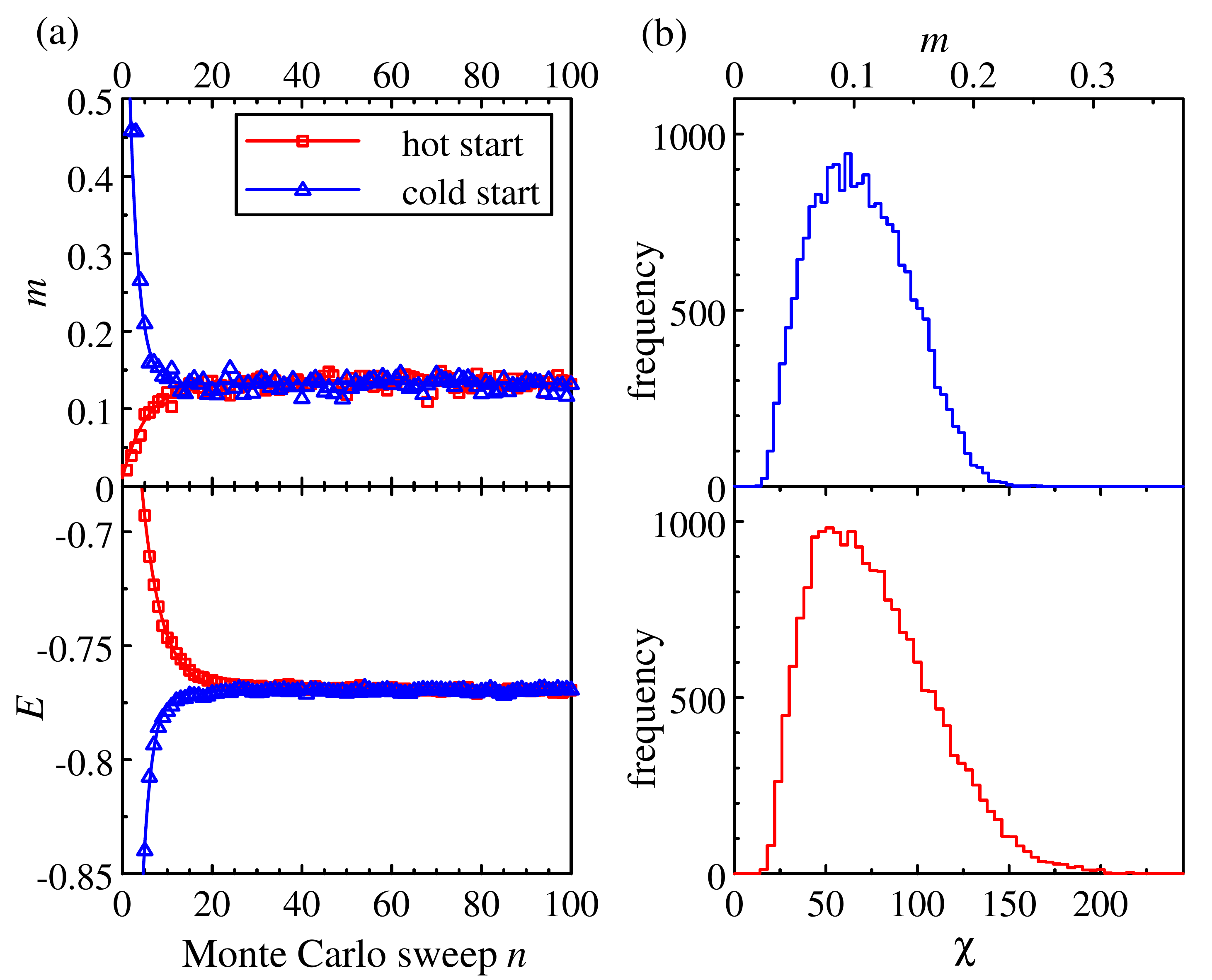}
\caption{(a) Equilibration of energy and order parameter for a single disorder realization
of size $L=100, L_\tau=452$,
dilution $p=1/3$, and temperature $T=T_c=1.577$. The solid lines are fits to
$E_n = E_{av} + a \exp(-n/t_{eq})$ (and analogously for $m$).
(b) Histograms of the order parameter $m$ and the order parameter susceptibility $\chi$
for $L=44, L_\tau=132$, $p=1/3$, $T=T_c=1.577$, using 20,000 disorder realizations.
To obtain accurate values for each individual sample, 4000 measurements per sample
were used in this calculation.}
\label{fig:equil_distrib}
\end{figure}

We perform simulations of the classical Hamiltonian (\ref{eq:S_Hcl}) by employing a combination of Wolff
cluster updates \cite{Wolff89S} and Metropolis single-spin updates \cite{MRRT53S} for dilutions
$p=0, 1/8, 1/5$, and $1/3$. Specifically, a full Monte Carlo sweep consists of a Metropolis sweep
followed by a Wolff sweep (a number of cluster flips such that the total number of flipped spins equals the
system size). The Wolff updates greatly reduce the critical slowing down, and the Metropolis updates help
equilibrate small disconnected clusters of lattice sites. The resulting equilibration and correlation times
are very short. This can be seen in Fig.\ \ref{fig:equil_distrib}(a) which shows the equilibration of the
energy $E$ and the order parameter $m$ for a ``worst-case example'', i.e., high dilution,
large system size, and a temperature right at criticality, $T=T_c$.
The figure demonstrates that the data for a hot start (random $\mathbf{S}$) and a cold start
(all $\mathbf{S}$ perfectly aligned) rapidly overlap. Fits of the energy to $E_n = E_{av} + a \exp(-n/t_{eq})$
(and analogously for the order parameter) give equilibration times $t_{eq}$ of about 5 sweeps or shorter,
depending on the quantity and initial conditions.

Due to the large computational effort required for simulating disordered systems, one must carefully choose
the number $N_S$ of disorder realizations (samples) and the number $N_M$ of measurements during the
simulation of each sample for optimal performance. Assuming statistical independence between
different measurements (quite possible with a cluster update), the variance $\sigma_T^2$
of the final result (thermodynamically and disorder averaged) for a particular observable
can be estimated as
\begin{equation}
\sigma_T^2 \approx (\sigma_S^2 + \sigma_M^2/N_M)/N_S
\end{equation}
where $\sigma_S^2$ is the disorder-induced variance between samples and $\sigma_M^2$ is the
variance of measurements within each sample \cite{BFMM98S,ZWNHV15S}. Because the numerical effort is
roughly proportional to $(N_M+N_E) N_S$ (where $N_E$ is the number of equilibration sweeps per sample),
the optimum value of $N_M$ is quite small.
We therefore employ a large number $N_S$ of disorder
realizations, ranging from 10\,000 to 20\,000,  and rather short runs of $N_M=500$
sweeps, with measurements taken after every sweep. The equilibration period for
each sample is $N_E=100$ Monte Carlo sweeps, many times longer than the longest observed equilibration
times $t_{eq}$.
The combination of short Monte Carlo runs and large sample numbers can lead to biases
in some observables, at least if the usual estimators are employed.
We have corrected these biases by means of improved estimators, as discussed,
e.g., in the appendix of Ref.\ \cite{ZWNHV15S}.
Moreover, for selected parameters, we have compared runs
using as little as 250 and as many as 4000 measurement sweeps per sample and confirmed that
they agree within their error bars.

To ascertain the importance of rare events, we compute the probability distributions
of key observables. Figure \ref{fig:equil_distrib}(b) shows histograms of the order parameter
and the order parameter susceptibility for dilution $p=1/3$ right at the critical temperature.
The distributions are moderately broad and do not feature long tails. This agrees with what is
expected based on the classification of disordered phase transitions \cite{VojtaSchmalian05S,*VojtaHoyos14S}.
The thermodynamic critical behavior of the superfluid-Mott glass transition is of conventional
power-law type \cite{Vojtaetal16S}, implying a finite-disorder fixed point.

As the disorder breaks the symmetry between space and imaginary time, we need to distinguish
the system sizes $L$ (in the space directions) and $L_\tau$ (in the imaginary-time direction).
Appropriate sample shapes can be found from the maxima of the Binder cumulant as described in Refs.\
\cite{GuoBhattHuse94S,RiegerYoung94S,SknepnekVojtaVojta04S,*VojtaSknepnek06S}. This method yields
combinations of $L$ and $L_\tau$ with constant scaling ratio $L_\tau/L^z$. To ensure that our results
are not affected by finite-size effects, we use large systems with linear sizes up to
$L=256$ and $L_\tau=512$ in the space and imaginary time directions. These sizes are much larger
than the correlation lengths (in the space and imaginary time directions)
of the studied excitations. For example, the smallest Higgs mass
(peak frequency) for the clean system shown in the inset of Fig.\ 1 of the main paper is
$m_H \approx 0.14$ corresponding
to a characteristic time of $2\pi/m_H \approx 45$, much smaller than the system size $L=L_\tau = 128$.
To gain further confidence, we have nonetheless confirmed that the results do not change for a
system of size $L=L_\tau = 256$.
In the presence of disorder, finite-size effects are even less of a problem because the Higgs mode
localizes. (The maximum of the spectral density in the main panel of Fig.\ 1 of the main paper is
at a frequency of about 1.25 corresponding to a characteristic time of about 5). We have confirmed this
by comparing the results for sizes between $L=68, L_\tau=256$ and $L=109, L_\tau=512$ for dilution
$p=1/3$.

To calculate the scalar susceptibility $\chi_{\rho\rho}$, we need to measure the local order parameter magnitude.
In a ``soft-spin'' model, one could simply use $|\mathbf{S}_{i,\tau}|$ for this purpose. However, in our
XY model, $|\mathbf{S}_{i,\tau}|$ is fixed at unity. We therefore define the local order-parameter magnitude
via an average over a small five-site cluster,
\begin{equation}
\rho(\mathbf{x}_i,\tau) = \frac 1 5  \bigg | \epsilon_i \mathbf{S}_{i,\tau} + \sum_j \epsilon_j \mathbf{S}_{j,\tau}   \bigg  |
\end{equation}
where the sum is over the four (space) neighbors of lattice site $i$.

\section*{S2. Scaling form of the scalar susceptibility}

Podolsky and Sachdev \cite{PodolskySachdev12S} derived a scaling form of the scalar susceptibility $\chi_{\rho\rho}$
at the clean superfluid-Mott insulator transition. We generalize this derivation to the Mott glass case by including
quenched (random-mass) disorder and a dynamical exponent $z$ different from unity. We start from a $d$-dimensional
quantum field theory for an $M$-component vector order parameter $\psi$; it is defined by the action
\begin{equation}
S= \int d^dx d\tau \left [ (\partial_{\mathbf x} \psi)^2 + (\partial_\tau \psi)^2 +(r + \delta r(\mathbf x) ) \psi^2 + u \psi^4 \right]~.
\label{eq:S_action}
\end{equation}
Here, $\delta r(\mathbf x)$ represents the quenched random mass disorder and $u$ is the standard quartic coefficient.
For $d=2$ and $M=2$, the quantum phase transition of this field theory is in the same universality class as the
superfluid-Mott glass transition of the Bose-Hubbard model (\ref{eq:S_Hubbard}). The corresponding free-energy density
is given by
\begin{equation}
f= - \frac 1 {\beta V} \ln Z = - \frac 1 {\beta V} \ln \int D[\psi] e^{-S}
\label{eq:S_free_energy}
\end{equation}
where $V$ is the system volume and $\beta$ the inverse temperature. We take two derivatives of $f$ w.r.t.\
the distance $r$ from criticality yielding
\begin{eqnarray}
\frac {\partial^2 f}{\partial r^2} &=& \frac 1 {\beta V} \int d^dx d\tau \int d^dx'd\tau'  \\
    && \times \left[ \langle \psi^2 (\mathbf x, \tau) \psi^2 (\mathbf x', \tau')\rangle - \langle \psi^2 (\mathbf x, \tau) \rangle \langle \psi^2 (\mathbf x', \tau')\rangle \right] ~. \nonumber
\label{eq:S_scalar}
\end{eqnarray}
This is the $\mathbf q =0$, $\omega_n =0$ Fourier component of the scalar susceptibility $\chi_{\rho\rho}$. (Actually, the expression yields the
correlation function of the square of the order parameter magnitude rather than the magnitude itself. However, as the magnitude has a nonzero average
at criticality, both these correlation functions have the same scaling behavior.)

The singular part of the free-energy density fulfills the homogeneity relation $f(r) = b^{-(d+z)} f(rb^{1/\nu})$ where $b$ is an arbitrary scale factor.
Taking two derivatives w.r.t.\ $r$ gives the scale dimension of $\chi_{\rho\rho}$ as $-(d+z) + 2/\nu$. This implies the homogeneity relation
\begin{equation}
\chi_{\rho\rho}(r,\mathbf q, \omega) = b^{-(d+z) + 2/\nu} \chi_{\rho\rho}(r b^{1/\nu},\mathbf q b, \omega b^z) ~.
\end{equation}
Setting $b=r^{-\nu}$ yields the scaling form
\begin{equation}
\chi_{\rho\rho}(r,\mathbf q, \omega) = r^{(d+z)\nu -2}  Y(\mathbf q r^{-\nu}, \omega r^{-\nu z})
\label{eq:S_scaling_r}
\end{equation}
or, equivalently
\begin{equation}
\chi_{\rho\rho}(r,\mathbf q, \omega) = \omega^{[(d+z)\nu -2]/(\nu z)}  X(\mathbf q r^{-\nu}, \omega r^{-\nu z})
\label{eq:S_chi_scaling}
\end{equation}
as given in Eq.\ (4) of the main text. Setting $z=1$  and $d=2$, we recover the result of
Podolsky and Sachdev for the clean superfluid-Mott insulator transition.

\begin{figure}
\includegraphics[width=8.5cm]{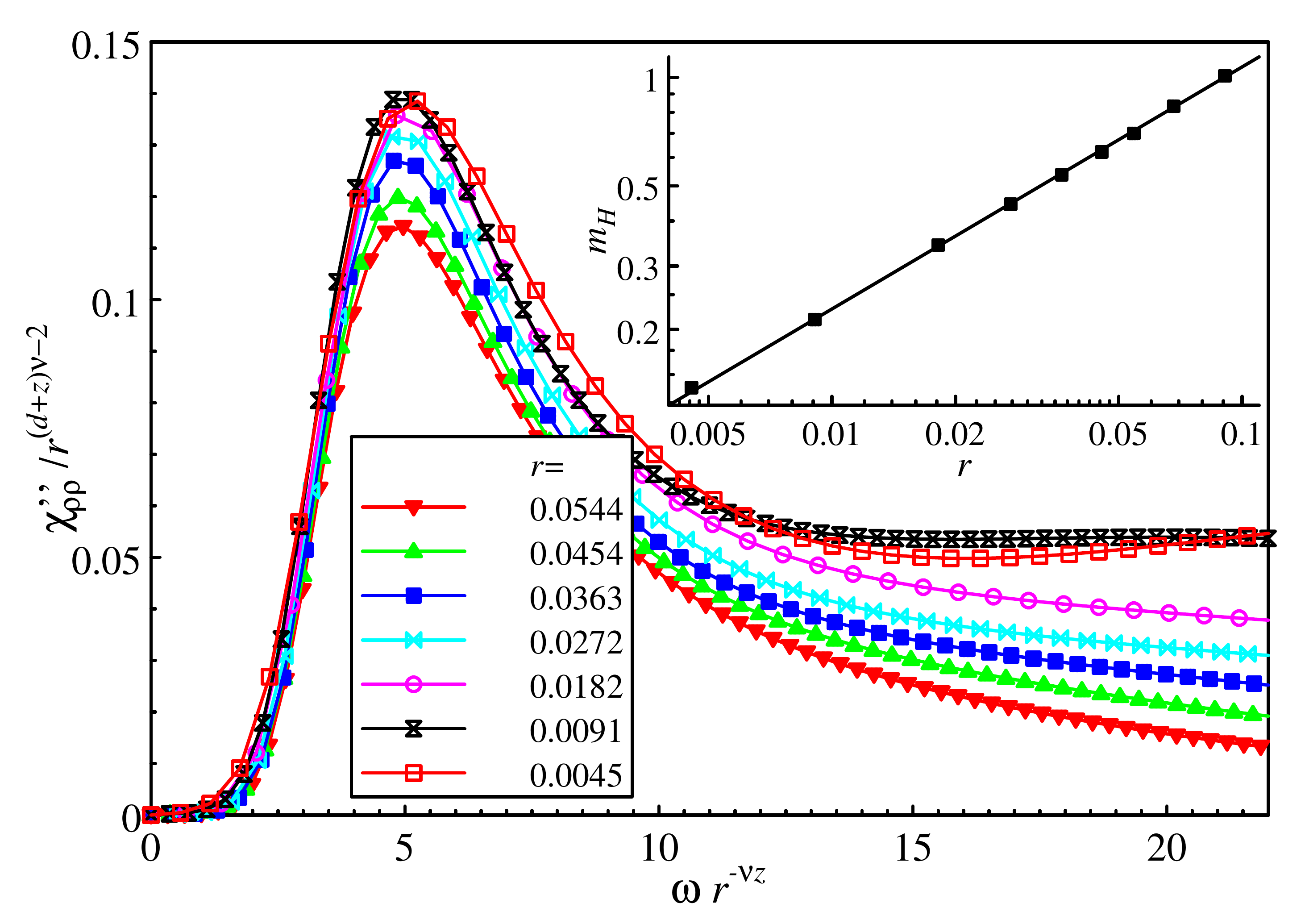}
\caption{Scaling plot of the spectral function $\chi_{\rho\rho}'' (r,\mathbf q=0, \omega)$ on the superfluid
side of the quantum phase transition in the undiluted case, $p=0$.
The results are averages over 2,000 samples of size $L=L_\tau=256$.
Inset: Energy of the Higgs peak in $\chi_{\rho\rho}''$, i.e., the Higgs mass $m_H$ vs.\ distance
from criticality $r$. The solid line is a fit of the expected power-law behavior $m_H \sim |r|^{\nu z}$
using the exponent values $\nu z=0.671$ of the 3d XY universality class \cite{CHPV06S}.
}
\label{fig:S_chi_scaling}
\end{figure}

As an illustration, Fig.\ \ref{fig:S_chi_scaling} presents a scaling plot of our Monte Carlo data for
the spectral function $\chi_{\rho\rho}'' (r,\mathbf q=0, \omega)$ in the undiluted case, $p=0$.
The figure shows that the low-energy part of $\chi_{\rho\rho}''$ fulfills the scaling form (\ref{eq:S_scaling_r})
in good approximation. For $r \lesssim 0.02$, the collapse of the Higgs peaks is nearly perfect. For larger $r$,
the peak positions continue to follow the expected power-law behavior (as is also demonstrated in the inset of the figure)
but the peak amplitudes show some deviations.
This can be attributed to uncertainties of the maximum-entropy method, as is discussed in the next section.

It is interesting to analyze the scale dimension of $\chi_{\rho\rho}$ or, equivalently, the power of $\omega$
in front of the scaling function $X$ in Eq.\ (\ref{eq:S_chi_scaling}). In a disordered system, the correlation length
exponent $\nu$ is known to fulfill the inequality $d\nu >2$ \cite{CCFS86S}. The exponent of $\omega$ thus fulfills the inequality
\begin{equation}
[(d+z)\nu -2]/(\nu z) > 1~.
\end{equation}
This positive exponent implies that, in the presence of disorder, the amplitude of the singular part of $\chi_{\rho\rho}$ is
strongly suppressed to zero as the quantum critical point is approached. Using the numerical values for $z$ and $\nu$ found in Ref.\
\cite{Vojtaetal16S}, the exponent takes the value $[(d+z)\nu -2]/(\nu z) = 1.18$ in our problem.

\section*{S3. Maximum-entropy method}

Within the Monte Carlo simulations, we compute the scalar susceptibility in imaginary time.
The real-frequency susceptibility is given by the analytic continuation
\begin{eqnarray}
\chi_{\rho\rho}(\mathbf q, \omega)  = \tilde \chi_{\rho\rho}(\mathbf q, i\omega_m \to \omega +i0^+)~
\label{eq:S_Wick}
\end{eqnarray}
from imaginary Matsubara frequencies $i \omega_m$ to real frequencies $\omega$.
This amounts to inverting the relation
\begin{equation}
\tilde \chi_{\rho\rho}(\mathbf q, i\omega_m) = \frac 1 \pi \int_0^\infty d\omega  \chi_{\rho\rho}''(\mathbf q, \omega) \frac{2\omega}{\omega_m^2 +\omega^2}
\label{eq:S_transformation}
\end{equation}
between the Matsubara susceptibility and the real-frequency spectral function $A(\mathbf q, \omega)=\chi_{\rho\rho}''(\mathbf q, \omega)$.
Unfortunately, the kernel of this transformation, $K(\omega_m, \omega)=(2/\pi)\omega/(\omega_m^2+\omega^2)$, is an ill-conditioned operator.
This renders the inversion extremely sensitive to the unavoidable noise in the numerical data.

To overcome this problem, we employ a version of the maximum-entropy method \cite{JarrellGubernatis96S}.
To find the spectral function $A$, we minimize (with respect to the spectral function $A$
that we wish to determine) the cost function
\begin{equation}
Q = \Delta - \alpha S~.
\end{equation}
The first term in $Q$, the error sum
\begin{equation}
\Delta = (\tilde \chi -K A)^T \Sigma  (\tilde \chi -K A)~,
\end{equation}
evaluates how well the spectral function fits the numerical data.
Here, $\tilde \chi$ represents the numerical data for the Matsubara susceptibility,
$K A$ is a shorthand for the transformation (\ref{eq:S_transformation}), and
$(\Sigma^{-1})_{mn} = \langle\tilde\chi(i\omega_m)\tilde\chi(i\omega_n)\rangle$ is the
covariance matrix in Matsubara space of the numerical
data.  The second term in $Q$ contains the entropy
\begin{equation}
S= - \sum_\omega A(\omega) \ln A(\omega)
\end{equation}
of the spectral function; it regularizes the inversion. The relative weights of the two terms in $Q$
is determined by the parameter $\alpha$ which we fix by a version of the L-curve method \cite{HansenOleary93S,BergeronTremblay16S}.

\begin{figure}
\includegraphics[width=\columnwidth]{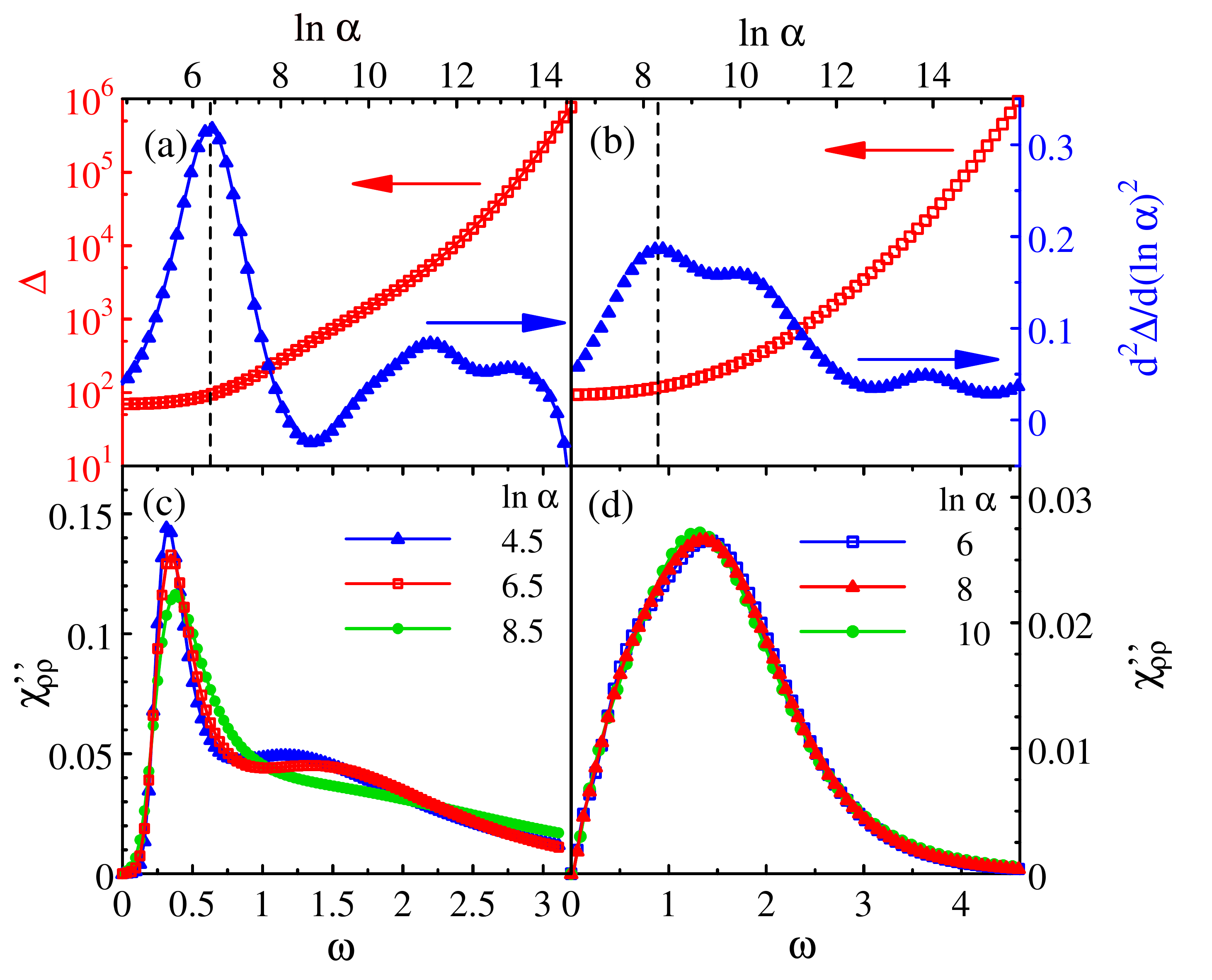}
\caption{Maximum-entropy method for the Higgs spectral function.
(a) Error sum $\Delta$ vs.\ $\ln \alpha$ for a clean ($p=0$) system with $L=L_\tau=256$ and
distance $r=-0.018$ from criticality. The optimal $\ln \alpha = 6.4$ is marked by a dashed line.
(b) $\Delta$ and $d^2 \Delta/d(\ln \alpha)^2$ vs.\ $\ln \alpha$ for dilution $p=1/3$, $L=68$, $L_\tau=256$
at $r=-0.049$. The optimal $\ln \alpha = 8.3$ is marked by a dashed line.
(c) Spectral density $\chi_{\rho\rho}''$ for $p=0$ and several values of $\ln \alpha$.
(d) Spectral density $\chi_{\rho\rho}''$ for $p=1/3$ and several values of $\ln \alpha$.}
\label{fig:maxent_test}
\end{figure}

The details of the maximum-entropy method, as applied to our data, are illustrated in
Fig.\ \ref{fig:maxent_test}.
Panel (a) of the figure shows the error sum $\Delta$ as a function of $\alpha$ for an example of a clean
system, and panel (b) does the same for a diluted system with $p=1/3$. Within the L-curve method, the optimal
$\alpha$ is determined by the maximum of the curvature $d^2 \Delta/d(\ln \alpha)^2$ which marks the crossover
from fitting the data (at larger $\alpha$) to fitting the Monte Carlo noise (at lower $\alpha$).
As a consistency check, we verify that the value of $\Delta$ at the optimal $\alpha$ approximately equals
the number of degrees of freedom, i.e., the number $L_\tau/2$ of independent Matsubara frequencies.

Panel (c) of Fig.\ \ref{fig:maxent_test} presents the resulting spectral densities for the clean example
for a range of $\alpha$ around the optimal value, and panel (d) does the same for the diluted system.
The sharp Higgs peak in the clean spectral density is affected by the value of $\alpha$, but only for
sizable deviations of $\alpha$ from its optimum value. In these cases the peak amplitude is more
sensitive than the peak frequency which changes by less than 10\%. In contrast, the broad ``hump''
in the spectral density of the diluted system remains essentially unchanged over a broad range of $\alpha$ values.

We have further tested the robustness of the maximum-entropy method by varying the ranges of
included real and Matsubara frequencies. As long as the included frequencies cover the main features
of the spectral density, this leads to small changes in $\chi_{\rho\rho}''$ of just a few percent.

We estimate the statistical error of the spectral density $\chi_{\rho\rho}''$ by means of an
ensemble method. We create an ensemble of artificial data sets from the Monte Carlo data for $\chi_{\rho\rho}(\mathbf q, i\omega_m)$ by adding Gaussian noise to the data points, with the
variance of the noise identical to the statistical uncertainties of the Monte Carlo data.
We then determine $\chi_{\rho\rho}''$ for each of the data sets by a separate maximum-entropy
calculation. A statistical analysis of all these results yields the error bars of $\chi_{\rho\rho}''$.
Applying this method to our data, we find that the statistical errors of $\chi_{\rho\rho}''$ are small,
about one symbol size in Figs.\ \ref{fig:maxent_test}(c) and (d).

\section*{S4. Inhomogeneous mean-field approach}

We start from the square-lattice Bose-Hubbard Hamiltonian (\ref{eq:S_Hubbard}) with large integer filling $\bar{n}$.
We truncate the local Hilbert space at each site $j$ to the three basis states $|-_j\rangle$, $|0_j\rangle$, and $|+_j\rangle$, corresponding to occupation numbers $n_j=\bar{n}-1$, $\bar{n}$, and $\bar{n}+1$, respectively.

We now perform a basis transformation in each local Hilbert space by introducing new basis states
\begin{widetext}
\begin{eqnarray}
	|\phi_{0j}\rangle &=&\cos{(\theta_j/2)} |0_j\rangle + \sin{(\theta_j/2)} \left(e^{i\eta_j} |+_j\rangle + e^{-i\eta_j} |-_j\rangle\right)/\sqrt{2} ~, \\
	|\phi_{Hj}\rangle &=& \sin{(\theta_j/2)} |0_j\rangle - \cos{(\theta_j/2)} \left(e^{i\eta_j}|+_j\rangle + e^{-i\eta_j}|-_j\rangle\right)/\sqrt{2} ~, \\
	|\phi_{Gj}\rangle &=& i (e^{i\eta_j}|+_j\rangle-e^{-i\eta_j}|-_j\rangle)/\sqrt{2}~.
\end{eqnarray}
\end{widetext}
The inhomogeneous mean-field theory is based on a product ansatz for the ground-state wave function, $|\Phi_0\rangle=\prod_{j} |\phi_{0j}\rangle$.
It interpolates between the Mott limit ($\theta_j=0$) and the superfluid limit ($\theta_j=\pi/2$). The local superfluid order parameter reads
$\langle a_j^\dagger \rangle \propto  \psi_j = \sin \theta_j e^{-i \eta_j}$.
The other two basis states, $|\phi_{Hj}\rangle$ and $|\phi_{Gj}\rangle$, correspond to changes of the order parameter amplitude and the order parameter phase, respectively, compared to the local ground state $|\phi_{0j}\rangle$.
The local variational parameters, i.e., the mixing angles $\theta_j$ and the phases $\eta_j$ are obtained by minimizing the ground state energy
\begin{eqnarray}
E_0 &=& \langle \Phi_0 | H | \Phi_0 \rangle \\
    &=& \frac 1 2 \sum_j U_j \sin^2 \frac {\theta_j} 2 -\sum_{\langle ij \rangle} \bar n J_{ij} \sin \theta_i \sin\theta_j \cos (\eta_i-\eta_j), \nonumber
\end{eqnarray}
leading to constant phases $\eta_j=\text{const}$ (which we set to zero) and mixing angles that fulfill the coupled mean-field equations
\begin{eqnarray}
	4\bar{n}\cos(\theta_i)\sum_{j} J_{ij} \sin(\theta_j) =  U_i \sin(\theta_i) \;.
\end{eqnarray}

To describe excitations on top of the mean-field solution, we introduce boson operators $b_{0j}^\dagger$, $b_{Hj}^\dagger$, and $b_{Gj}^\dagger$ that create the local basis states $|\phi_{Hj}\rangle$, $|\phi_{Hj}\rangle$ and $|\phi_{Gj}\rangle$ out of the fictitious vacuum state. They fulfill the local constraint
$b_{0j}^\dagger  b_{0 j} + b_{H j}^\dagger  b_{H j} + b_{G j}^\dagger  b_{G j} = 1$. We now rewrite the Bose-Hubbard Hamiltonian in terms of the $b$ bosons
and use the constraint to eliminate $b_{0j}$. To quadratic order in $b$, the excitation modes decouple and the Hamiltonian becomes the sum of a Higgs
Hamiltonian and a Goldstone Hamiltonian, $H=E_0+H_H+H_G$, with
\begin{widetext}
\begin{eqnarray}
	H_{H}=&&-\sum_{\langle ij \rangle} \bar n J_{ij} \cos{\theta_i} \cos{\theta_j} (b_{Hi}^\dagger + b_{H i}) (b_{Hj}^\dagger + b_{Hj}) + \sum_{i}\varpi_{H,i} b_{H j}^\dagger  b_{H j} \;, \\
	H_{G}=&&- \sum_{\langle ij \rangle} \bar nJ_{ij} \cos{(\theta_i/2)} \cos{(\theta_j/2)} (b_{Gi}^\dagger + b_{G i}) (b_{Gj}^\dagger + b_{G j})  + \sum_{i}\varpi_{G,i} b_{G j}^\dagger  b_{G j} \; .
\end{eqnarray}
\end{widetext}
Each Hamiltonian describes a set of coupled harmonic oscillators with local frequencies $\varpi_{Hi}= U_i\cos(\theta_i)/2+2\zeta_i$ and $\varpi_{Gi}= U_i\cos^2(\theta_i/2)/2+\zeta_i$ where $\zeta_i=\sin(\theta_i)\sum_{j} \bar n J_{ij}\sin(\theta_j)$.

The Hamiltonians $H_H$ and $H_G$ can each be diagonalized by Bogoliubov transformations ($\alpha=G,H$)
\begin{eqnarray}
b_{\alpha j} = \sum_k ( u_{\alpha jk} d_{\alpha k} + v_{\alpha jk}^\ast d_{\alpha k}^\dagger )
\end{eqnarray}
where the $d$ bosons correspond to the collective mode eigenstates of our disordered system.
The transformation coefficients $u$ and $v$ turn out to be real, they are given by
\begin{eqnarray}
u_{\alpha jk} &=& \frac 1 2\mathcal{V}_{\alpha jk} \left(\sqrt{\varpi_{\alpha j}/\nu_{\alpha k}} + \sqrt{\nu_{\alpha k}/\varpi_{\alpha j}} \right)~, \\
v_{\alpha jk} &=& \frac 1 2 \mathcal{V}_{\alpha jk} \left(\sqrt{\varpi_{\alpha j}/\nu_{\alpha k}} - \sqrt{\nu_{\alpha k}/\varpi_{\alpha j}} \right)~.
\end{eqnarray}
The matrix $\mathcal{V}_{\alpha jk}$ contains the eigenvectors (as columns) of the collective-mode eigenvalue problem
\begin{eqnarray}
	\sum_j X_{\alpha ij} \mathcal{V}_{\alpha jk} = \nu_{\alpha k}^2 \mathcal{V}_{\alpha ik}
\end{eqnarray}
where $\nu_{\alpha k}$ are the nonnegative excitation eigenfrequencies (energies). The coupling matrix $X$ reads
\begin{eqnarray}
 	{X}_{Gij} &=& \varpi^2_{Gi} \delta_{ij} -2 \cos{(\theta_i/2)} \cos{(\theta_j/2)} \bar n J_{ij} \sqrt{\varpi_{Gi}\varpi_{Gj}}  \nonumber\\
 	{X}_{Hij} &=& \varpi^2_{Hi} \delta_{ij} - 2 \cos{\theta_i} \cos{\theta_j} \bar n J_{ij} \sqrt{\varpi_{Hi}\varpi_{Hj}}
\end{eqnarray}
for the Goldstone and Higgs mode, respectively.

In terms of the $d$ bosons, $H_H$ and $H_G$ are diagonal,
\begin{equation}
 H_H= \sum_i \nu_{H i} d_{H i}^\dagger d_{H i}~, \qquad  H_G= \sum_i \nu_{G i} d_{G i}^\dagger d_{G i}
\end{equation}

Using this mean-field approach, we analyze systems with site dilutions $p=0, 1/8, 1/5$, and $1/3$.
We consider square lattices with up to $256^2$ sites as well as quasi-onedimensional strips of up
to $128 \times 10^6$ sites.

\section*{S5. Localization properties of the Higgs and Goldstone excitations}

To study the localization properties of the Bogoliubov states, we analyze both the participation number $P$
and the effective fractal dimension $\tau_2$ of the eigenstates. The inverse participation number $P^{-1}$
of state number $k$ is given by  \cite{VojtaM13S}
\begin{equation}
P^{-1}(k)=\sum_j (|u_{\alpha jk}|^2 - |v_{\alpha jk}|^2)^2 = \sum_j |\mathcal{V}_{\alpha jk}|^4~.
\end{equation}
To define the fractal dimension, we divide the system into boxes of linear size $l$.
We define a measure
\begin{equation}
\mu_b = \sum_{j \in b} (|u_{\alpha jk}|^2 - |v_{\alpha jk}|^2) = \sum_{j \in b} |\mathcal{V}_{\alpha jk}|^2
\end{equation}
characterizing the probability of state $k$ in box $b$ as well as its second moment
\begin{equation}
P_{l}^{-1}(k) = \sum_b \mu_b^2~.
\end{equation}
Note that we recover the participation number for box size $l=1$, i.e., $P(k)=P_{1}(k)$.
The corresponding fractal dimension reads
\begin{equation}
\tau_{2}(k) =  \ln P_{l}(k)  / \ln (L/l) ~.
\label{eq:tau2}
\end{equation}
The asymptotic value of $\tau_2$ is obtained in the limit $L/l \to \infty$.

\begin{figure}
\includegraphics[width=8.5cm]{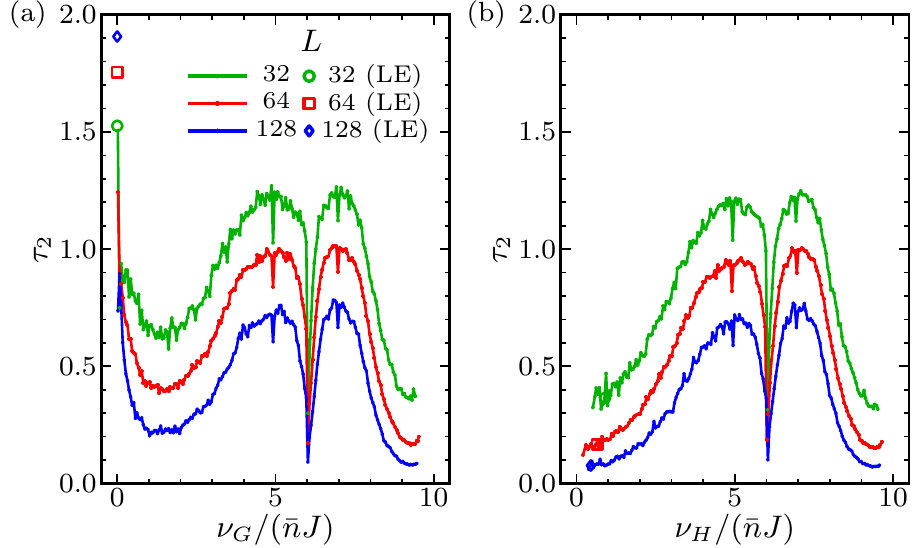}
\caption{Generalized fractal dimension $\tau_2$ of Goldstone (a) and Higgs (b) excitations vs.\ excitation energy $\nu$
for $U/(\bar n J)=12$, dilution $p=1/3$ and several system sizes $L$ with $L/l=8$. The solid lines represent averages of
$\tau_2$ over small energy windows (width 0.1) and 100 to 400 disorder configurations, depending on $L$.
The values of $\tau_2$ of the lowest-energy
excitation (averaged over all disorder configurations) are shown as open symbols.}
\label{fig:S_tau_2}
\end{figure}

Figure \ref{fig:S_tau_2} illustrates the energy dependence of the effective dimension $\tau_2$ for
Higgs and Goldstone excitations for dilution $p=1/3$ at $U/(\bar n J)=12$, slightly on the superfluid side of the transition.
For the Higgs mode, $\tau_2$ decreases with system size $L$ for all energies indicating that the entire band
is localized. The same holds for the Goldstone mode at any nonzero excitation energy. In contrast, the lowest
energy Goldstone mode ($\nu_G=0$) shows the opposite scaling behavior. $\tau_2$ increases towards 2 with increasing $L$,
indicating an extended state. Note that the Higgs and Goldstone modes show almost identical behavior
for larger excitation energies, $\nu \gtrsim 3$, reflecting that they are still almost degenerate close
to the quantum phase transition. The sharp features at energies around $\nu=6$ are the result of the discrete
character of the site dilution used to implement the disorder.

\begin{figure}
\includegraphics[width=8.5cm]{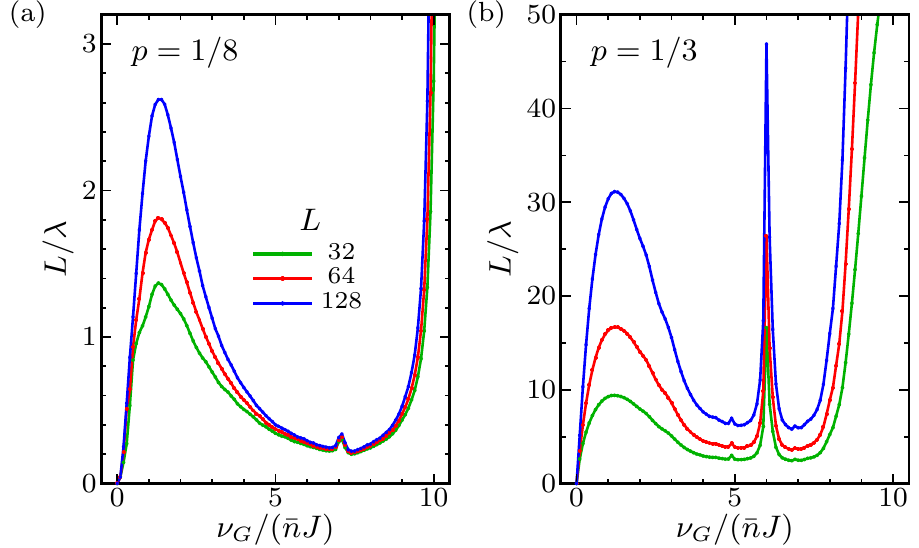}
\caption{Scaled inverse localization length $L/\lambda$ of the Goldstone excitations vs.\ excitation energy
$\nu_G$, calculated using the iterative Green's function method on strips of $L \times 10^6$ sites (the data
are averages over 12 strips).
For dilution $p=1/3$ the data are taken at $U/(\bar n J)=12)$, for dilution $p=1/8$
the data are for $U/(\bar n J)=14$.}
\label{fig:S_lambda}
\end{figure}

In addition to the multifractal analysis of the eigenstates, we also apply the iterative Green's function method
\cite{MacKinnonKramer83S,MacKinnon85S}
to quasi-onedimensional strips. Within this methods, the localization length $\lambda$
is calculated from the decay of
the Green's function between the two ends of the strip. Normalizing $\lambda$ by the strip width $L$ yields a dimensionless
quantity suitable for finite-size scaling. Figure \ref{fig:S_lambda} presents the energy dependence
of the Goldstone mode localization length for systems slightly in the superfluid phase for dilutions $p=1/3$ and $1/8$.
Specifically, it shows the scaled inverse localization length $L/\lambda$
as function of the excitation energy $\nu_G$  and several strip widths $L$.
The data for both weak and strong dilution display the same qualitative behavior. At all nonzero energies, $L/\lambda$
increases with increasing strip width indicating that the Goldstone mode is localized. However, $L/\lambda$ decreases rapidly
as the energy $\nu_G$ approaches zero, and for $\nu_G=0$ the inverse localization length vanishes for all strip width.
These results confirm the findings of the multifractal analysis of the eigenstates above.

\section*{S6. Analytic expression for the lowest Goldstone excitation}

According to Goldstone's theorem, the lowest eigenstate of the Goldstone Hamiltonian $H_G$ must have zero energy, $\nu_{G0}= 0$,
in the superfluid phase because the superfluid ground state spontaneously breaks the $U(1)$ order-parameter symmetry.
For this state, the corresponding eigenvalue problem
\begin{equation}
\sum_{j} X_{Gij} \mathcal{V}_{G j0} = \nu_{G0}^2 \mathcal{V}_{G i0}=0
\label{S_eq_GS}
\end{equation}
simplifies to a system of linear equations. A non-trivial solution of this system is given by
\begin{eqnarray}
	\mathcal{V}_{G j0} = \varUpsilon \frac{\sin(\theta_j/2)} {\sqrt{\varpi_{Gj}}}
\label{eq:HandySolution}
\end{eqnarray}
as can be easily checked by inserting it back into the system (\ref{S_eq_GS}). Here,  $\varUpsilon$ is a normalization constant.
Thus, the lowest Goldstone eigenstate depends on the order parameter $\sin (\theta_j)$ and local interactions (via $\varpi_{Gj}$) only.

The denominator in (\ref{eq:HandySolution}) is bounded from both below and above. Specifically, in our site-diluted system,
$\varpi_{Gj} \ge U /4 $ and $\varpi_{Gj} \le U /2 + 4\bar n J$.  Consequently, the localization character of
$\mathcal{V}_{G j0}$ agrees with that of the order parameter.

Let us now assume the the system features a nonzero macroscopic order parameter $\psi$, i.e., an average order parameter
that is nonzero in the thermodynamic limit). This implies either a more-or-less homogeneous superfluid or at lest a nonzero
density of superfluid puddles. According to Eq.\ (\ref{eq:HandySolution}), this means that the wave function
of the lowest Goldstone excitation is nonzero on a finite fraction of the sites, i.e., it is extended.

In the Mott phase, where $\sin(\theta_j)$ vanishes on all sites,  the state (\ref{eq:HandySolution}) is not normalizable, indicating the absence of a zero-energy mode. It is also interesting to note that  $\sin(\theta_j)=0$ in the Mott phase implies that the disorder in the coupling matrix $X_{Gij}$ is produced by $U_i$ and $J_{ij}$ only. The disorder is thus uncorrelated in space guaranteeing that all states are localized in two dimensions.

\end{document}